\documentclass[aps,pra,preprint,groupedaddress]{revtex4-2}
\usepackage[dvips]{graphicx}
\usepackage{amsmath,amssymb}
\usepackage{mathrsfs}
\usepackage{float}

\begin{document}

%\preprint{}

\title{Self-modulation of nonlinear light 
in vacuum enhanced by orbital angular momentum}


\author{Akitaka Watanabe}
\affiliation{Solution Technology Dept. Tokyo Gas Co.,Ltd., 
1-5-20 Kaigan, Minato-ku, Tokyo,105-8527 Japan}

\author{Kazunori Shibata}
\email{shibata-ka@ile.osaka-u.ac.jp}
\affiliation{Institute of Laser Engineering, Osaka University, 2-6 Yamada-Oka, Suita, Osaka 565-0871 Japan}

\date{\today}

\begin{abstract}
Nonlinear optical effects in vacuum have been investigated as a means to verify quantum electrodynamics in a region of low photon energy. By considering nonlinear electromagnetic waves in a three-dimensional cylindrical cavity, we report that the orbital angular momentum of light strongly affects self-modulations in a long timescale. The variation in optical phase is shown to enhance the vacuum nonlinearity. Moreover, we demonstrate the time evolution of the energy transfer between cavity modes and of the phase shift, paving new possibility for verification experiments. 
\end{abstract}

%\keywords{}

\maketitle

\section{Introduction}

Nonlinear electromagnetism in vacuum treats the interactions between electromagnetic fields. 
Such a nonlinearity is supposed to 
stem from virtual electron-positron pairs, i.e., 
the Heisenberg-Euler theory\cite{HE1,Heisenberg1936}. 
For the theory and 
other models\cite{doi:10.1098-rspa.1934.0059,Plebanski1970}, the 
lowest-order nonlinear electromagnetic Lagrangian density is 
characterized by two small 
parameters $C_{2,0}$ and $C_{0,2}$.
However, these values have not yet been identified experimentally.

The nonlinearity in vacuum is considered to 
cause peculiar phenomena such as 
magnetic monopoles\cite{PhysRevD.63.044005}. 
Moreover, a correction brought by 
the nonlinearity is studied in astrophysics, e.g., 
on the radiation from pulsars or magnetars\cite{Heyl_2005,Shakeri_2017,10.1093/mnras/stw2798} 
and 
a state of black holes\cite{AYONBEATO199925,PhysRevD.63.044005}.

Towards the detection of vacuum nonlinearity, 
one proposal is to employ a high-intensity and 
ultra short-pulse laser \cite{RRP-S145}. 
Another method is to utilize a mirror and low- or modest-intensity laser, 
such as a waveguide \cite{PhysRevLett.87.171801}, 
ring laser\cite{Denisov2001}, and cavity. 
Several experiments have been performed, e.g., the 
PVLAS(Polarizzazione del Vuoto con LASer)\cite{PhysRevD.90.092003,Della_Valle_2013}, BMV(Bir\'efringence Magn\'etique du Vide)\cite{Cadene2014}, and OVAL(Observing VAcuum with Laser)\cite{Fan2017} experiments, which aimed to detect vacuum birefringence.

A cavity system is capable of retaining light 
in much longer time than short-pulse lasers. 
In a cavity, a resonant increase of nonlinear correction with time has been theoretically studied 
\cite{Shibata2020,Shibata2021EPJD,PhysRevA.70.013808,PhysRevA.105.013508}. 
Recently, an appearance of large self-modulation 
in a long timescale has been reported in 
one- and two- dimensional 
cavities\cite{PhysRevA.104.063513,Shibata2022EPJD}.
The self-modulation in the 
long timescale can become 
comparable to classical fields. 
For verification experiments,
it is worthwhile to reveal characteristics 
of a large self-modulation 
in a long timescale in three-dimensional cavity.

In this study, we consider a three-dimensional cylindrical cavity, 
with a modest-intensity laser and static magnetic field. 
We elucidate that the orbital angular momentum of light completely changes the behavior of self-modulation. 
 As an example of self-modulation, 
energy transfer 
among the cavity modes is demonstrated. 

This manuscript is organized as follows. 
The basic notation, considered system, and classical 
electromagnetic fields are explained in the next section. 
The resonant term in a linear approximation is given in Sec. 3. 
In Sec. 4, the differential equations in Eq. (\ref{eq:bibun_6hensu}) 
are derived. They are the key equations which describe 
the large self-modulation in 
a longer timescale.  The solution of Eq. (\ref{eq:bibun_6hensu}) is 
given in Sec. 5. Section 6 is dedicated to demonstrate how the 
angular momentum of light changes the self-modulation. 
An experimental perspective is stated in Sec. 7. 
Final remarks are given in the last section.

\section{Notation, system, and classical term}

We normalize electromagnetic fields by the electric constant 
$\varepsilon_0$ and magnetic constant $\mu_0$. 
The electric field $\boldsymbol{E}$ is multiplied by 
$\varepsilon_0^{1/2}$ and 
the magnetic flux density $\boldsymbol{B}$ 
is divided by $\mu_0^{1/2}$, respectively. 
Quantum electrodynamics predicts that 
the vacuum yields a nonlinear effect on electromagnetic fields 
via virtual electron-positron pairs \cite{HE1,Heisenberg1936,PhysRev.82.664}. 
By using two Lorentz invariants 
$F=\boldsymbol{E}^2-\boldsymbol{B}^2$ and 
$G=\boldsymbol{E}\cdot\boldsymbol{B}$, 
the lowest-order nonlinear electromagnetic Lagrangian is given by 
\begin{equation}
\mathscr{L}=\frac{1}{2}F+C_{2,0}F^2+C_{0,2}G^2, 
\end{equation}
where $C_{2,0}$ and $C_{0,2}$ are the nonlinear parameters. 
The values in the Heisenberg-Euler theory are 
$C_{2,0}=\hbar e^4/(360\pi^2\varepsilon_0^2m_e^4c^7)$ 
and $C_{0,2}=7C_{2,0}$, 
respectively
\cite{PhysRev.82.664,PhysRevD.93.093020,PhysRevD.98.056001}, 
where 
$\hbar$ the Planck's constant divided by $2\pi$, 
$e$ the elementary charge, 
$m_e$ the electron mass, and 
$c$ the speed of light.

The electric flux density $\boldsymbol{D}$ 
and magnetic field $\boldsymbol{H}$ are derived by the 
partial derivatives of $\mathscr{L}$ with respect to 
$\boldsymbol{E}$ and $-\boldsymbol{B}$, 
respectively. The vacuum nonlinearity appears 
in these constitutive equations. 
The nonlinear Maxwell's equations are 
\begin{equation}
\begin{split}
&\nabla\times\boldsymbol{E}+\frac{1}{c}\frac{\partial\boldsymbol{B}}{\partial t}=\boldsymbol{0}, \\
&\nabla\cdot\boldsymbol{B}=0, \\
&\nabla\times\boldsymbol{H}-\frac{1}{c}\frac{\partial\boldsymbol{D}}{\partial t}=\boldsymbol{0}, \\
&\nabla\cdot\boldsymbol{D}=0. \\
\end{split}
\label{eq:teigi_MX}
\end{equation}

\subsection{System}

\begin{figure}
\includegraphics[width=100mm]{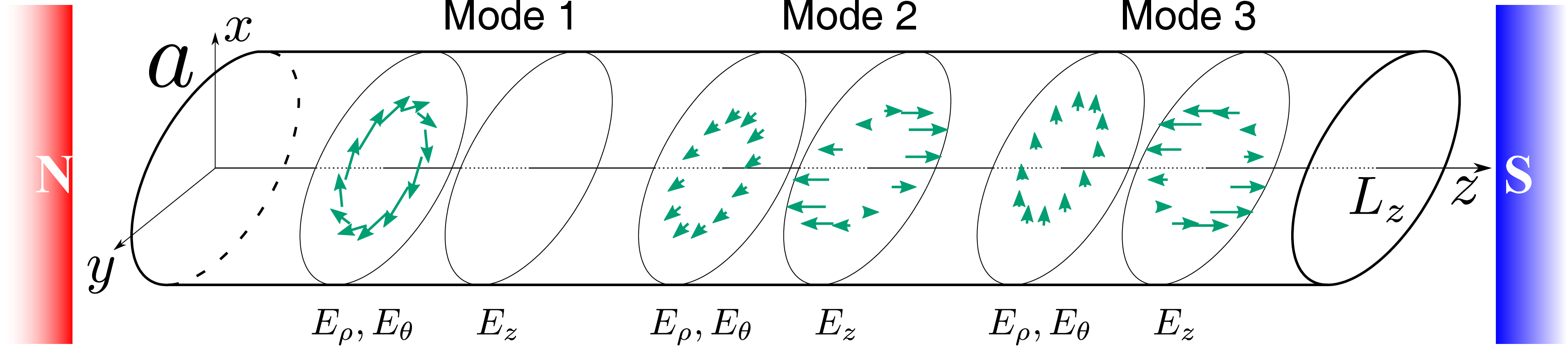}
\caption{
The cylindrical cavity system with a perfect mirror. 
The electric fields for three eigenmodes are 
illustrated in the $xy$ plane at $\rho=a/2$. 
The external magnetic flux density is symbolically 
shown by the N and S poles. 
\label{fig_kyoushinki}}
\end{figure}

The whole system of cylindrical cavity is 
depicted in Fig. \ref{fig_kyoushinki}. 
The radius and height of the cylindrical cavity 
are given by $a$ and $L_z$, respectively. 
The cavity mirror is supposed to be 
a perfect conductor. Then, the boundary conditions are 
determined. This system is dissipation-free, i.e., 
no energy loss from reflection.

\subsection{Classical term}

The total nonlinear electromagnetic field is divided into 
two parts. One is well-known ``classical term'' 
which 
satisfies the linear classical Maxwell's equations. 
As the classical term, we 
consider three eigen cavity modes, i.e., standing waves. 
They have the same frequency $\omega$. 
We call them mode 1, 2, and 3, respectively. 
The wavenumber components in radial and axial directions are given by 
$k_{\rho}=\lambda_{1j}/a$ and $k_z=n\pi/L_z$, where 
$n\in\mathbb{N}$ and $\lambda_{1j}$ denotes the 
$j$-th positive zero of the Bessel function $J_1$. 
The wavenumber is given by 
$k=(k_{\rho}^2+k_z^2)^{1/2}=\omega/c$.
We notate $T=\omega t$. 
In cylindrical coordinate system 
$(\rho,\theta,z)$, the electric fields for each eigenmode 
$\boldsymbol{E}_{c1}, \boldsymbol{E}_{c2}
$, and $\boldsymbol{E}_{c3}$ are 
given as 
\begin{equation}
\begin{split}
E_{c1\rho}&=0, \\
E_{c1\theta}&=\frac{1}{\sqrt{2}}J_1(k_{\rho}\rho)
\sin{k_zz}\left(f_c\cos{T}+f_s\sin{T}\right), \\
E_{c1z}&=0, \\
\end{split}
\label{eq:mode1E}
\end{equation}
\begin{equation}
\begin{split}
E_{c2\rho}&=-\frac{k_z}{2k}
\left[J_0(k_{\rho}\rho)-J_2(k_{\rho}\rho)\right]
\sin{\theta}\sin{k_zz}\left(g_c\cos{T}+g_s\sin{T}\right), \\
E_{c2\theta}&=-\frac{k_z}{2k}
\left[J_0(k_{\rho}\rho)+J_2(k_{\rho}\rho)\right]
\cos{\theta}\sin{k_zz}\left(g_c\cos{T}+g_s\sin{T}\right), \\
E_{c2z}&=\frac{k_{\rho}}{k} J_1(k_{\rho}\rho)
\sin{\theta}\cos{k_zz}\left(g_c\cos{T}+g_s\sin{T}\right), \\
\end{split}
\label{eq:mode2E}
\end{equation}
\begin{equation}
\begin{split}
E_{c3\rho}&=-\frac{k_z}{2k}
\left[J_0(k_{\rho}\rho)-J_2(k_{\rho}\rho)\right]
\cos{\theta}\sin{k_zz}\left(h_c\cos{T}+h_s\sin{T}\right), \\
E_{c3\theta}&=\frac{k_z}{2k}
\left[J_0(k_{\rho}\rho)+J_2(k_{\rho}\rho)\right]
\sin{\theta}\sin{k_zz}\left(h_c\cos{T}+h_s\sin{T}\right), \\
E_{c3z}&=\frac{k_{\rho}}{k}J_1(k_{\rho}\rho)
\cos{\theta}\cos{k_zz}\left(h_c\cos{T}+h_s\sin{T}\right), \\ 
\end{split}
\label{eq:mode3E}
\end{equation}
where $f_c, f_s, g_c, g_s, h_c,$ and $h_s$ are amplitudes. 
These magnitudes are modest as not to break the mirror. 
We also consider a constant static 
magnetic flux density $B_s$ 
in the $z$ direction, but not in the $x$ and $y$ directions 
because of the compatibility of Maxwell's equations 
and boundary conditions\cite{Shibata_2022}.

The classical orbital angular momentum 
is not zero only in the $z$ component, i.e., 
\begin{equation}\label{eq:koten_Y}
\frac{1}{c}\int \rho (E_zB_{\rho}-E_{\rho}B_z)\text{d}V=
-\frac{\pi a^2L_z}{4\omega}
(g_ch_s-g_sh_c)J_0(\lambda_{1j})^2. 
\end{equation}
We simply refer to the $z$ component as ``angular momentum''.

\section{Resonant increase in short timescale}

In the following sections, we calculate 
the ``corrective term'' which is the difference 
of total nonlinear electromagnetic fields and 
classical term. 

In a short timescale, e.g., $t \approx 2\pi/\omega$, 
the corrective term is small enough compared to the classical term. 
A perturbative linear approximation can be applied. 
The polarization 
$\boldsymbol{P}=\boldsymbol{D}-\boldsymbol{E}$ and 
magnetization 
$\boldsymbol{M}=\boldsymbol{B}-\boldsymbol{H}$ 
of vacuum are approximated only by the classical term. 
They are notated as $\boldsymbol{P}_c$ and 
$\boldsymbol{M}_c$, respectively.  
We express the corresponding corrective term 
as $\boldsymbol{E}^{(0)}_n$ and $\boldsymbol{B}^{(0)}_n$. 
They satisfy the following equations: 
\begin{equation}
\begin{split}
&\nabla\times\boldsymbol{E}^{(0)}_n
+k\dot{\boldsymbol{B}}{}^{(0)}_n=\boldsymbol{0}, \\
&\nabla\cdot\boldsymbol{B}^{(0)}_n=0, \\
&\nabla\times\boldsymbol{B}^{(0)}_n
-k\dot{\boldsymbol{E}}{}^{(0)}_n
=k\dot{\boldsymbol{P}}_c+\nabla\times\boldsymbol{M}_c, \\
&\nabla\cdot\boldsymbol{E}^{(0)}_n=
-\nabla\cdot\boldsymbol{P}_c, \\
\end{split}
\label{eq:maxwell_senkei}
\end{equation}
where an overdot expresses 
the differentiation with respect to $T$. 
A part of the solution is a resonant term 
that increases with time, i.e., the secular term. 
For example,  
$E_{\theta}$ component of the resonant term 
is given as 
\begin{equation}\label{eq:reso_mode1_2_3}
\begin{split}
E_{\theta}^{(reso)}=&
\frac{1}{\sqrt{2}}
\left(D_{1c}\cos{T}+D_{1s}\sin{T}\right)T
J_1(k_{\rho}\rho)\sin{k_zz}\\
%%%%%%%%%%%%%%%%%%%%%%%%%%
&-\frac{k_z}{2k}
\left(D_{2c}\cos{T}+D_{2s}\sin{T}\right)T
\left[J_0(k_{\rho}\rho)+J_2(k_{\rho}\rho)\right]
\cos{\theta}\sin{k_zz}\\
%%%%%%%%%%%%%%%%%%%%%%%%%%
&+\frac{k_z}{2k}
\left(D_{3c}\cos{T}+D_{3s}\sin{T}\right)T
\left[J_0(k_{\rho}\rho)+J_2(k_{\rho}\rho)\right]
\sin{\theta}\sin{k_zz}. \\
\end{split}
\end{equation}
It should be emphasized that the resonant term 
has the same spatial distribution as the classical mode. 
This feature is the same for the other components. 

The coefficients 
$D_{1c},D_{1s},D_{2c},D_{2s},D_{3c}$, and $D_{3s}$ 
are given as follows.
Let $\Delta=k_{\rho}^2/k^2$ and 
\begin{equation}\label{def_X1-X7}
\begin{split}
S_1&=
\frac{1}{4\lambda_{1j}^2J_0(\lambda_{1j})^2}C_{2,0}
\left[
8(1-\Delta+\Delta^2)I_+
+7\Delta^2I_-
\right], \\
%%%%%%%%%%%%%%%%%%%%%%%%%%%%%%%%%%%%%%%%%
S_2&=\frac{1}{48\lambda_{1j}^2J_0(\lambda_{1j})^2}
\biggr\{
24C_{2,0}\left[
-4\Delta(1-\Delta)I_+
+(4-4\Delta-\Delta^2)I_-
\right] \\
&\quad+C_{0,2}\left[
8I_+
-(8-9\Delta^2)I_-
\right]\biggr\}, \\
%%%%%%%%%%%%%%%%%%%%%%%%%%%%%%%%%%%%%%%%%
S_3&=\frac{1}{24\lambda_{1j}^2J_0(\lambda_{1j})^2}
\biggl\{8C_{2,0}\left[
2(1+2\Delta-2\Delta^2)I_+
+(10-10\Delta+\Delta^2)I_-\right] \\
&\quad-C_{0,2}\left[
4I_+
-(4-3\Delta^2)I_-
\right]\biggr\}, \\
%%%%%%%%%%%%%%%%%%%%%%%%%%%%%%%%%%%%%%%%%
S_4&=\frac{1}{24\lambda_{1j}^2J_0(\lambda_{1j})^2}C_{2,0}
\left[72(1-\Delta+\Delta^2)I_+
-5(16-16\Delta+9\Delta^2)I_-
\right], \\
%%%%%%%%%%%%%%%%%%%%%%%%%%%%%%%%%%%%%%%%%
S_5&=\frac{1}{12\lambda_{1j}^2J_0(\lambda_{1j})^2}
\biggl\{
C_{2,0}\left[
-24(1-\Delta+\Delta^2)I_+
+(16-64\Delta+15\Delta^2)I_-\right] \\
&\quad +8C_{0,2}(1-\Delta)I_-
\biggr\}, \\
\end{split}
\end{equation}
where 
\begin{equation}\label{Integral_I+I-}
\begin{split}
&I_+=\int^{\lambda_{1j}}_0xJ_1(x)^4\text{d}x,\\
&I_-=\int^{\lambda_{1j}}_0\frac{1}{x}J_1(x)^4\text{d}x, \\
\end{split}
\end{equation}
and let $\mathscr{A}_1=f_c^2+f_s^2, 
\mathscr{A}_{23}=g_c^2+g_s^2+h_c^2+h_s^2$, 
the resonant coefficients are given by 
\begin{equation}\label{eq:kyoumeikeisu_teisu}
\begin{split}
%%%%%%%%%%%%%%%%%
D_{1c}=&
-\left(S_1\mathscr{A}_1+S_2\mathscr{A}_{23}
+4\frac{k_{\rho}^2}{k^2}C_{2,0}B_s^2\right)f_s
-S_3\left[\left(f_cg_c+f_sg_s\right)g_s
+\left(f_ch_c+f_sh_s\right)h_s\right], \\
%%%%%%%%%%%%%%%%%
D_{1s}=&
\left(S_1\mathscr{A}_1+S_2\mathscr{A}_{23}
+4\frac{k_{\rho}^2}{k^2}C_{2,0}B_s^2\right)f_c
+S_3\left[\left(f_cg_c+f_sg_s\right)g_c
+\left(f_ch_c+f_sh_s\right)h_c\right], \\
%%%%%%%%%%%%%%%%%
D_{2c}=&
-\left(S_2\mathscr{A}_1
+S_4\mathscr{A}_{23}+\frac{k_{\rho}^2}{k^2}C_{0,2}B_s^2\right)g_s
-S_3\left(f_cg_c+f_sg_s\right)f_s
+S_5\left(g_ch_s-g_sh_c\right)h_c, \\
%%%%%%%%%%%%%%%%%
D_{2s}=&
\left(S_2\mathscr{A}_1
+S_4\mathscr{A}_{23}+\frac{k_{\rho}^2}{k^2}C_{0,2}B_s^2\right)g_c
+S_3\left(f_cg_c+f_sg_s\right)f_c
+S_5\left(g_ch_s-g_sh_c\right)h_s, \\
%%%%%%%%%%%%%%%%%
D_{3c}=&
-\left(S_2\mathscr{A}_1
+S_4\mathscr{A}_{23}
+\frac{k_{\rho}^2}{k^2}C_{0,2}B_s^2\right)h_s
-S_3\left(f_ch_c+f_sh_s\right)f_s
-S_5\left(g_ch_s-g_sh_c\right)g_c, \\
%%%%%%%%%%%%%%%%%
D_{3s}=&
\left(S_2\mathscr{A}_1
+S_4\mathscr{A}_{23}
+\frac{k_{\rho}^2}{k^2}C_{0,2}B_s^2\right)h_c
+S_3\left(f_ch_c+f_sh_s\right)f_c
-S_5\left(g_ch_s-g_sh_c\right)g_s. \\
\end{split}
\end{equation}

Other terms such as high harmonics are 
always much smaller than the classical term 
unless the particular cavity size $a/L_z$ 
\cite{PhysRevA.70.013808}. 
We do not treat such special situation here.

The corrective term within the 
linear approximation can be written in the 
form of ``$\boldsymbol{E}^{(0)}_n=\boldsymbol{E}^{(reso)}+$non-resonant terms''. 
The vacuum nonlinearity becomes more detectable 
as the corrective term increases. However, 
such a time evolution eventually violates the 
linear approximation. 
The applicable time limit cannot be determined within the 
linear approximation, although it must be much longer 
than the short timescale $2\pi/\omega$. 
Then, we proceed beyond the linear approximation. 
We analyze the behavior of 
the leading term and show how 
the angular momentum plays a role. This is the main theme 
of this study.

\section{Leading term beyond linear approximation}

Hereinafter, we refer to the term ``long timescale'' 
only to intend that we are considering beyond the linear approximation. 
 
Recall that the resonant term 
has the same spatial distribution as the classical term. 
This means that 
the leading term of the total 
nonlinear electromagnetic waves shall 
have the same spatial distributions,  
even in a long timescale 
beyond the applicable limit of the linear approximation.

Therefore, handling the leading term can be accomplished 
by regarding the amplitudes 
$f_c, f_s, g_c, g_s, h_c,$ and $h_s$ ``time dependent''. 
These are assumed to change slowly enough compared to 
$2\pi/\omega$, in a viewpoint of multiscale analysis. 
In their Taylor expansions, 
the zeroth-order term corresponds to the classical 
amplitude. The first-order term corresponds to
the resonant coefficients. 
Hence, by using Eq. (\ref{eq:reso_mode1_2_3}), 
nonlinear simultaneous differential equations 
for slowly varying 
$f_c, f_s, g_c, g_s, h_c,$ and $h_s$ are derived as 
\begin{equation}\label{eq:bibun_6hensu}
\begin{split}
&\dot{f}_c=D_{1c}, \ \ \dot{f}_s=D_{1s}, \\
%%%%%%%%%%%%%%%%%%%%%%%
&\dot{g}_c=D_{2c}, \ \ \dot{g}_s=D_{2s}, \\
%%%%%%%%%%%%%%%%%%%%%%%
&\dot{h}_c=D_{3c}, \ \ \dot{h}_s=D_{3s}. \\
\end{split}
\end{equation}
Their initial values are set to 
\begin{equation}\label{def_6hensu_syoki}
\begin{split}
&f_c(0)=0, \ \ f_s(0)=A_{(1)}, \\
&g_c(0)=A_{(2)}\sin{\varphi_{(2)}}, \ \ 
g_s(0)=A_{(2)}\cos{\varphi_{(2)}}, \\
&h_c(0)=A_{(3)}\sin{\varphi_{(3)}}, \ \ 
h_s(0)=A_{(3)}\cos{\varphi_{(3)}}, \\ 
\end{split}
\end{equation}
where $A_{(1)},A_{(2)},$ and $A_{(3)}$ are the 
classical amplitudes of modes $1,2,$ and $3$, and 
$\varphi_{(2)}$ and 
$\varphi_{(3)}$ are relative phases of 
modes $2$ and $3$ to mode $1$, respectively.

Equation (\ref{eq:bibun_6hensu}) yields 
two conserved quantities as 
\begin{equation}\label{eq:hozon_6hensu}
\begin{split}
&X=f_c^2+f_s^2+g_c^2+g_s^2+h_c^2+h_s^2
=A_{(1)}^2+A_{(2)}^2+A_{(3)}^2>0, \\
&Y=g_ch_s-g_sh_c=A_{(2)}A_{(3)}\sin(\varphi_{(2)}-\varphi_{(3)}), \\
\end{split}
\end{equation}
where $X$ corresponds to the 
classical total energy and $Y$ the angular momentum. 
For this reason, we conclude that Eq. (\ref{eq:bibun_6hensu}) is reasonable. 
Note that $C_{2,0}X$ 
is dimensionless and practically, $C_{2,0}X\ll1$. 

To solve Eq. (\ref{eq:bibun_6hensu}), 
we introduce an auxiliary function $\alpha$ 
which expresses the energy ratio of the mode 1. 
It is solved in the appendix. In particular, we derive 
another conserved quantity $Z$.

\section{Solution of Eq. (\ref{eq:bibun_6hensu})}\label{sec:keisan_alp}

The solution of Eq. (\ref{eq:bibun_6hensu}) can be written down by 
using obtained $\alpha$. 
Please refer to the appendix for 
$Z, Q_1(\alpha), Q_2(\alpha), c_1, c_2, \mathscr{X}, 
\tilde{\mathscr{X}}, \mathscr{Y}, \tilde{\mathscr{Y}}, 
\xi, \mathfrak{q}_{1-}, \mathfrak{q}_{2-}, \tilde{\mathfrak{s}},$ 
and $Y_{max}$.

First, we show $f_c$ and $f_s$. If $\alpha=0$ at a time, 
it is always zero. Consequently, $f_c=0$ and $f_s=0$. 
Otherwise, by using 
\begin{equation}\label{eq:phi_neq0}
\Psi=\frac{1}{2}\left(S_1-S_4
\right)X
\int^T_0\alpha(\tau)\text{d}\tau
%%%%%%%%%%%%%%%%%%%%%
+\left(S_4X+
\frac{k_{\rho}^2}{k^2}C_{0,2}B_s^2\right)T
-\frac{Z}{2}\int^T_0\frac{1}{\alpha(\tau)}\text{d}\tau, 
\end{equation}
these are calculated as 
\begin{equation}\label{eq:fg_zyuubun}
f_c=-\sqrt{X\alpha}\sin\Psi, \ \ 
f_s=\sqrt{X\alpha}\cos\Psi. 
\end{equation}

Next, we show $g_c, g_s, h_c,$ and $h_s$. 
In the case of $c_1>0$ and $Y\neq0$, 
we obtain 
\begin{equation}\label{eq:lmpq}
\begin{split}
&g_c=-\sqrt{\frac{X}{c_1\alpha}}
\left( \sqrt{Q_1}\cos\Theta_1 \sin\Psi
+\sqrt{Q_2}\cos\Theta_2 \cos\Psi \right), \\
%%%%%%%%%%%%%%%%%%%%%%%%%%%%%%%%%
&g_s=\sqrt{\frac{X}{c_1\alpha}}
\left( \sqrt{Q_1}\cos\Theta_1 \cos\Psi
-\sqrt{Q_2}\cos\Theta_2 \sin\Psi \right), \\
%%%%%%%%%%%%%%%%%%%%%%%%%%%%%%%%%
&h_c=-\sqrt{\frac{X}{c_1\alpha}}
\left( \sqrt{Q_1}\sin\Theta_1 \sin\Psi
+\sqrt{Q_2}\sin\Theta_2 \cos\Psi \right), \\
%%%%%%%%%%%%%%%%%%%%%%%%%%%%%%%%%
&h_s=\sqrt{\frac{X}{c_1\alpha}}
\left( \sqrt{Q_1}\sin\Theta_1 \cos\Psi
-\sqrt{Q_2}\sin\Theta_2 \sin\Psi \right), \\
\end{split}
\end{equation}
where 
\begin{equation}
\begin{split}
\Theta_1&=c_1\frac{Y}{X}\int^T_0
\frac{(\xi+\mathscr{X})\alpha(\tau)-Z}
{Q_1(\alpha(\tau))}
\text{d}\tau+\tilde{\mathscr{Y}}T+\theta_1, \\
\Theta_2&=c_1\frac{Y}{X}\int^T_0
\frac{(\xi+\tilde{\mathscr{X}})\alpha(\tau)-Z}
{Q_2(\alpha(\tau))}
\text{d}\tau
+\mathscr{Y}T+\theta_2, \\
\end{split}
\end{equation}
and the constants 
$\theta_1, \theta_2 \in[0,2\pi)$ are 
uniquely determined by the initial values. 
These functions for $c_1=0$ or $Y=0$ 
 are obtained comparatively easily.

Finally, we have figured out 
the leading part of the nonlinear electromagnetic wave 
in the long timescale.

\section{Angular momentum affects self-modulation}

Here we demonstrate that 
the angular momentum $Y$ changes 
the energy transfer between the three modes. 
Equation (\ref{eq:hozon_6hensu}) shows that 
$Y$ depends on the intensity of mode 2 and 3 through 
$A_{(2)}$ and $A_{(3)}$, and the relative phase 
$\varphi_{(2)}-\varphi_{(3)}$. The intensity obviously affects 
the self-modulation. Hence, we focus on how the 
initial phases change the self-modulation. 
We fix $A_{(2)}$ and $A_{(3)}$ 
and vary only $\varphi_{(2)}$ and $\varphi_{(3)}$, 
to change $Y$ while keeping $X$ and $Z$ constant.

The numerical parameters are given 
as follows. 
We choose $j=3$ and $k_{\rho}^2/k^2=0.9$. 
Then, $c_1,c_2$, and $\mathscr{X}$ are 
$c_1/(C_{2,0}X)\approx 0.020$, 
$c_2/(C_{2,0}X)\approx 0.182$, and 
$\mathscr{X}/(C_{2,0}X)\approx 0.097$. 
The concrete values for $a,L_z,$ and $n$ 
are unnecessary.  
We further set $\xi/(C_{2,0}X)=0.03$ 
and $Z/(C_{2,0}X)=0.06$, corresponding to the first
line in Table \ref{table_a_shindou} at $Y=0$, 
where $\mathfrak{q}_{2-}\approx 0.275$ 
and $\mathfrak{q}_{1-}\approx 0.301$. 
Then, the maximum angular momentum is calculated as 
$Y_{max}/X\approx 0.356$ 
and the corresponding intersection is 
$\tilde{\mathfrak{s}}\approx 0.287$. 
We set $\alpha(0)=\tilde{\mathfrak{s}}$ 
to make sure that $Y$ can reach $Y_{max}$. 
The initial energy ratios are 
$A_{(1)}^2/X=\tilde{\mathfrak{s}} \approx 0.287$ and 
$A_{(2)}^2/X=A_{(3)}^2/X=(1-\tilde{\mathfrak{s}})/2
\approx 0.356$.

\begin{figure}
\includegraphics[width=100mm]{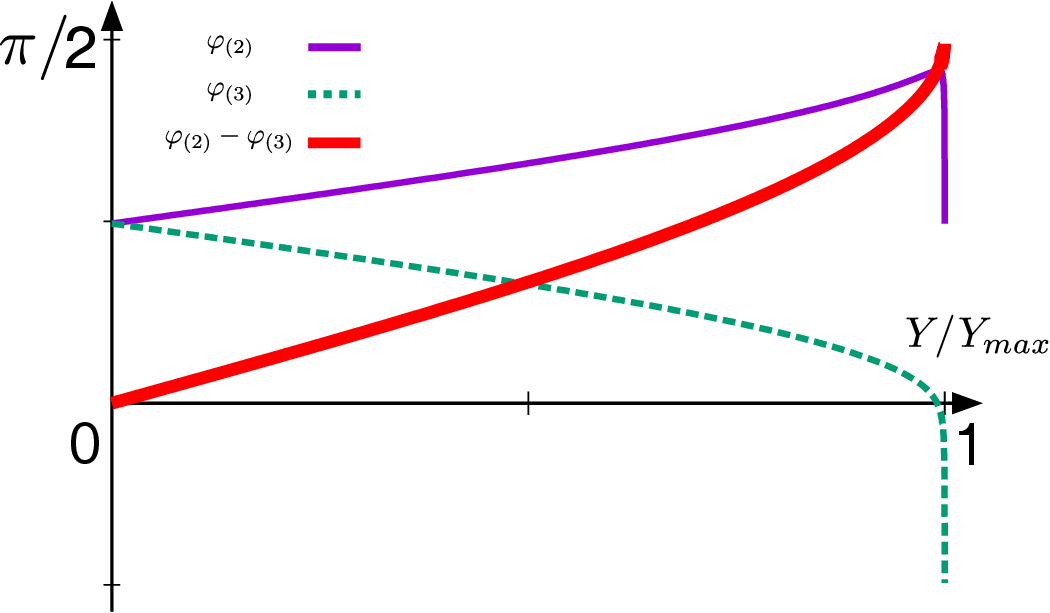}
\caption{
The initial phases $\varphi_{(2)}$ and $\varphi_{(3)}$ 
as functions of $Y$. 
The conserved quantity $Z$ introduced in 
Eq. (\ref{eq:betagamma2zyo}) is same for every 
pair of $\varphi_{(2)}$ and $\varphi_{(3)}$. 
The red bold curve expresses 
the relative phase 
$\varphi_{(2)}-\varphi_{(3)}$, satisfying 
$Y\propto \sin(\varphi_{(2)}-\varphi_{(3)})$. 
\label{fig:isou}}
\end{figure}

Equation (\ref{eq:betagamma2zyo}) indicates that 
$\varphi_{(2)}$ and $\varphi_{(3)}$ are mutually dependent for the 
fixed $Z$. 
Figure \ref{fig:isou} shows the 
relationship between $Y$ and the initial phases 
which keep $Z$ constant.

\begin{figure}
\includegraphics[width=120mm]{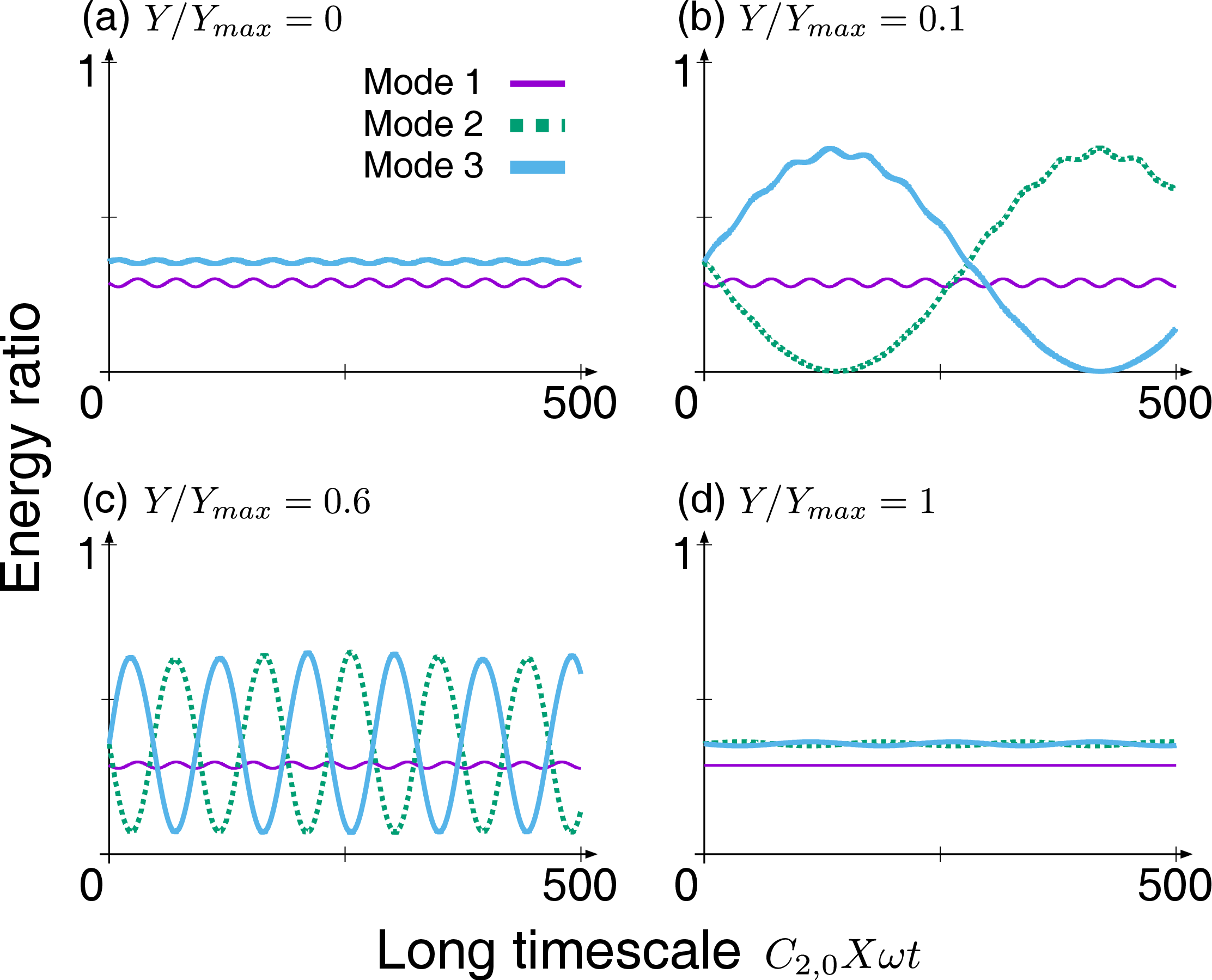}
\caption{
Time evolutions of the energy ratios 
in the long timescale for 
(a) $Y/Y_{max}=0$, (b) $Y/Y_{max}=0.1$, 
(c) $Y/Y_{max}=0.6$, and (d) $Y/Y_{max}=1$. 
\label{fig:ene}}
\end{figure}

Figure \ref{fig:ene}
shows the time evolutions of energy ratio of each mode 
for $Y/Y_{max}=0,0.1,0.6,1$. 
The horizontal 
axes are the long timescale $C_{2,0}XT$. Since $C_{2,0}X\ll1$, 
appearing oscillations are extremely slower than the 
one cycle of $2\pi/\omega$. 
As in Fig. \ref{fig:ene}(a), energy ratios of 
the modes 2 and 3 agree at $Y=0$. As $|Y|$ increases, 
the time evolutions of modes 2 and 3 have different 
periodicity against mode 1, shown in Figs. \ref{fig:ene}(b,c). 
The maximum energy ratio of modes 2 or 3 reaches 
up to about $0.722$ in Fig. \ref{fig:ene}(b). 
It is about twice the initial ratio $0.356$. 
For the maximum angular momentum 
$Y/Y_{max}=1$ in Fig. \ref{fig:ene}(d), the energy ratio of mode 1 
remains in the initial value as is 
$\alpha=\tilde{\mathfrak{s}}$. 
The energy ratios of modes 2 and 3 
are shifted in the half period of each other.

As demonstrated in Fig. \ref{fig:ene}, 
the self-modulation in a long timescale largely depends 
on the value of angular momentum. It is worth emphasizing 
that the self-modulation strongly depends on the initial phases, 
not only on the amplitudes, i.e., intensity.

It will be challenging to observe these behaviors 
in a current experiment. 
In a realistic cavity, light energy is eventually lost. 
If we let $t_{max}$ be the lifetime, 
it is typically several milliseconds for 
a tabletop cavity. 
For the visible light ($\omega \approx 
10^{15}$ s$^{-1}$) with intensity of 
$10^6$ W/cm$^2$, we obtain 
$C_{2,0}X\omega t_{max} \approx 10^{-16}$. 
Thus, we consider another example in the next section 
for a future experiment.

\section{Experimental perspective}\label{sec_EP}

In this example, the external magnetic field is more 
important than the angular momentum for 
a rapid emergence of the vacuum nonlinearity.

We suppose $A_{(1)}=0$, i.e., 
the initial state does not contain the mode 1. 
Then, Eq. (\ref{eq:bibun_6hensu}) is easily solved as 
\begin{equation}
\begin{split}
g_c&=-A_{(2)}\sin(\epsilon T-\varphi_{(2)})\cos S_5YT
-A_{(3)}\sin(\epsilon T-\varphi_{(3)})\sin S_5YT, \\
%%%%%%%%%%%%%%%%%%%%%%%%%%%%%
g_s&=A_{(2)}\cos(\epsilon T-\varphi_{(2)})\cos S_5YT
+A_{(3)}\cos(\epsilon T-\varphi_{(3)})\sin S_5YT, \\
%%%%%%%%%%%%%%%%%%%%%%%%%%%%%
h_c&=A_{(2)}\sin(\epsilon T-\varphi_{(2)})\sin S_5YT
-A_{(3)}\sin(\epsilon T-\varphi_{(3)})\cos S_5YT, \\
%%%%%%%%%%%%%%%%%%%%%%%%%%%%%
h_s&=-A_{(2)}\cos(\epsilon T-\varphi_{(2)})\sin S_5YT
+A_{(3)}\cos(\epsilon T-\varphi_{(3)})\cos S_5YT, \\
\end{split}
\end{equation}
where $\epsilon=S_4X+(k_{\rho}^2/k^2)C_{0,2}B_s^2$.

The electric field at $\theta=0$ and $z=0$ is approximated by 
\begin{equation}
E_z(\rho,0,0) \approx \frac{k_{\rho}}{k}J_1(k_{\rho}\rho)
\sqrt{h_c^2+h_s^2}\sin(T+\Theta), \\
\end{equation}
where the ``phase'' $\Theta$ is given by 
$\sin \Theta=h_c/\sqrt{h_c^2+h_s^2}$ and 
$\cos \Theta=h_s/\sqrt{h_c^2+h_s^2}$. 
The phase slowly varies with time due to 
the vacuum nonlinearity. 
The phase shift $\Delta \Theta=\Theta(t)-\Theta(0)$ 
can be controlled by the external magnetic field $B_s$. 
If both $|\epsilon \omega t_{max}|$ and 
$|S_5Y \omega t_{max}|$ are 
much smaller than unity, we obtain 
\begin{equation}
\Delta \Theta \approx 
-\left[S_5A_{(2)}^2\sin^2(\varphi_{(2)}-\varphi_{(3)})+
S_4X+\frac{k_{\rho}^2}{k^2}C_{0,2}B_s^2
\right]\omega t. \\
\end{equation}
In particular, if the external magnetic field is much 
stronger than the light intensity $I$, i.e., $B_s^2\gg (\mu_0/c)I$ (in SI units), we can evaluate the maximum phase shift by 
\begin{equation}
|\Delta \Theta| \approx 10^{-9}B_s^2t_{max}, \\
\end{equation}
where $B_s$ is in SI units and 
we applied $\omega=10^{15}$ s$^{-1}$.

The magnitude of static magnetic field is currently 
several tens tesla\cite{Majkic_2020}. 
$B_s \approx 10^2 $ T will be achieved in the near future. 
How can we evaluate $t_{max}$? 
Suppose the vacuum nonlinearity is detectable 
if 1\% of incident light remains. The mirror is 
supposed to be a supermirror with a reflectivity of 
99.9999\%. Let $L$ be the typical length between the mirrors. 
Then, we can evaluate $t_{max}$ by $0.999999^{ct_{max}/L}=0.01$. 
If the cavity size is huge as LIGO\cite{PhysRevLett.116.061102-2}, 
$L \approx 10^3$ m yields $t \approx 15$ s. 
Finally, $|\Delta \Theta| \approx 10^{-4}$ rad.

This estimation is somewhat artificial, as the original 
calculation is performed for a closed cavity 
with perfect mirrors. Moreover, 
it is not realistic to impose a uniform magnetic field 
for the whole cavity of kilometer size.

\begin{figure}
\includegraphics[width=120mm]{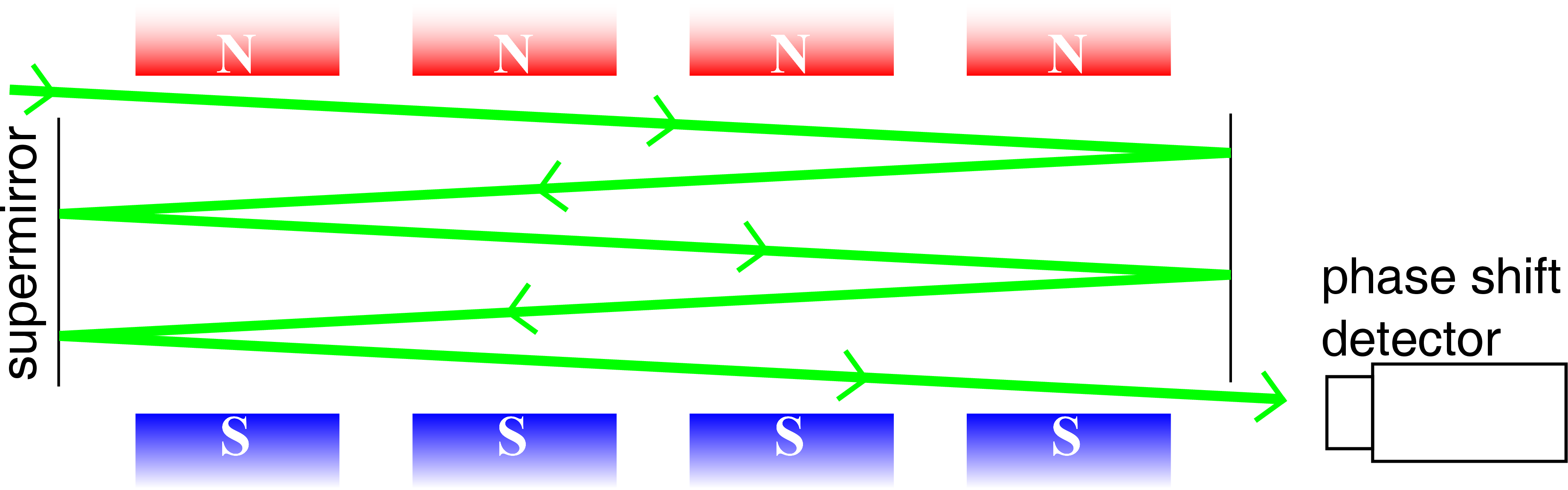}
\caption{
A successor system for an experimental observation 
of the vacuum nonlinearity. The incident light is 
depicted by a green zigzag line. It is 
modulated by the external magnets. 
The phase shift will be observed in the detector. 
\label{fig:opencavity}}
\end{figure}

A system in Fig. \ref{fig:opencavity} will be a next step. 
An open cavity is composed of supermirrors. 
The external magnetic field is imposed perpendicular to the 
cavity. Multiple magnets cover the whole cavity. 
The light is irradiated with a shallow angle. 
The phase shift of outgoing light is measured. 
The calculation of this system can be done 
by the extended FDTD method\cite{PhysRevA.104.063513}.

\section{Final remarks}

We have analyzed the nonlinear electromagnetic wave in 
a cylindrical cavity, especially in the viewpoint of 
angular momentum. 
Equation (\ref{eq:bibun_6hensu}) 
are derived by the secular term in 
Eq. (\ref{eq:reso_mode1_2_3}). 
The differential equations describe the self-modulation 
in the long timescale. 
We have demonstrated how the 
angular momentum changes the self-modulation. In particular, 
it has been elucidated that the energy transfer 
strongly depends on the initial phases. 
We would like to refer to the contribution 
of external field $B_s$. A large nonlinear 
behavior emerges sooner as the value becomes larger, 
as shown in Sec. \ref{sec_EP} and 
the previous studies 
\cite{PhysRevA.104.063513,Shibata2022EPJD}.

A third harmonic wave can 
increase resonantly if the cavity size $a/L_z$ 
satisfies a specific condition \cite{PhysRevA.70.013808}. 
To broaden a chance of verification experiment of the 
vacuum nonlinearity, it will be useful to analyze the 
high harmonics and other sum and difference 
frequencies, as performed in 
a rectangular cavity\cite{PhysRevA.105.013508}.

For a future verification experiment, 
we have considered another example where the external 
magnetic field plays an important role in this case. 
It will be worth pursuing a large open cavity system 
with supermirrors and external magnetic field.

The revealed behavior of the nonlinear electromagnetic wave 
in the cavity and the analysis method we performed 
will be of great merit for verifying the vacuum nonlinearity.

\begin{acknowledgments}
The authors thank to Dr. Nakai, Dr. Mima, and Dr. Seto 
for discussion. 
The authors quite appreciate Dr. J. Gabayno 
and Mr. Juadines for checking 
the logical consistency of the text. 
\end{acknowledgments}

% Specify following sections are appendices. Use \appendix* if there only one appendix.
\appendix*

\section{Solution method of Eq. (\ref{eq:bibun_6hensu})}

In this appendix, 
an auxiliary function $\alpha$ is introduced and solved. 
Its typical behavior is also shown. During the process, 
we obtain another conserved quantity $Z$.

\subsection{Reduction of functions}

There are 6 unknown functions in Eq. (\ref{eq:bibun_6hensu}). 
By introducing following 5 functions as  
\begin{equation}\label{eq:def_5hensu}
\begin{split}
\alpha&=\frac{1}{X}\left(f_c^2+f_s^2\right), \\
%%%%%%%%%%%%%%
\beta_2&=\frac{1}{X}\left(f_cg_c+f_sg_s\right), \ \ 
\beta_3=\frac{1}{X}\left(f_ch_c+f_sh_s\right), \\
%%%%%%%%%%%%%%
\gamma_2&=\frac{1}{X}\left(f_cg_s-f_sg_c\right), \ \ 
\gamma_3=\frac{1}{X}\left(f_ch_s-f_sh_c\right), \\
\end{split}
\end{equation}
reduced equations which contain 
only the 5 unknown functions are obtained:  
\begin{equation}\label{eq:bibun_5hensu}
\begin{split}
&\dot{\alpha}=c_1\left(\beta_2\gamma_2+\beta_3\gamma_3\right), \\
&\dot{\beta}_2=\left(c_2\alpha-\xi-\mathscr{X}\right)\gamma_2-\mathscr{Y}\beta_3, \\
&\dot{\beta}_3=\left(c_2\alpha-\xi-\mathscr{X}\right)\gamma_3+\mathscr{Y}\beta_2, \\
&\dot{\gamma}_2=-\left(\tilde{c}_2\alpha-\xi-\tilde{\mathscr{X}}\right)\beta_2-\tilde{\mathscr{Y}}\gamma_3, \\
&\dot{\gamma}_3=-\left(\tilde{c}_2\alpha-\xi-\tilde{\mathscr{X}}\right)\beta_3+\tilde{\mathscr{Y}}\gamma_2, \\
\end{split}
\end{equation}
where 
\begin{equation}\label{eq:para_keisan}
\begin{split}
&c_1=-2S_3X, \\
&c_2=\left(S_1-2S_2+S_4\right)X, 
\ \ 
\tilde{c}_2=c_2+c_1, \\ 
&\mathscr{X}=\left(S_4-S_2\right)X, \ \ 
\tilde{\mathscr{X}}=\mathscr{X}+\frac{1}{2}c_1, \\
&\mathscr{Y}=\left(S_3-S_5\right)Y, \ \ 
\tilde{\mathscr{Y}}=\mathscr{Y}+\frac{Y}{X}c_1, \\
&\xi=-\frac{k_{\rho}^2}{k^2}\left(4C_{2,0}-C_{0,2}\right)B_s^2. \\
\end{split}
\end{equation}
In particular, $\alpha \in [0,1]$ expresses 
the energy ratio of the mode 1.

The pair of second and third lines in 
Eq. (\ref{eq:bibun_5hensu}) yields a first integral. 
By defining
\begin{equation}
\begin{split}
&Q_1(\alpha)=c_2\alpha^2-2\left(\xi+\mathscr{X}\right)\alpha+Z, \\
&Q_2(\alpha)=c_1\alpha(1-\alpha)-Q_1(\alpha), \\
\end{split}
\end{equation}
where $Z$ is another conserved quantity, 
we obtain  
\begin{equation}
\begin{split}
c_1\left(\beta_2^2+\beta_3^2\right)&=Q_1(\alpha), \\
c_1\left(\gamma_2^2+\gamma_3^2\right)&=Q_2(\alpha). \\
\end{split}
\label{eq:betagamma2zyo}
\end{equation}
The magnitude of $Z$ is of the order of 
$C_{2,0}X$. This parameter is purely mathematical.

\subsection{Constraints on parameters}

We clarify the magnitude relations among 
the parameters. 
As for the two integrals 
$I_+$ and $I_-$ in Eq. (\ref{Integral_I+I-}), 
$I_+$ diverges and 
$I_-$ converges to $1/\pi^2$
\cite{Book_TISP,Book_HTF2}, as $j \to \infty$. 
Therefore, following inequalities hold for $j\ge3$: 
\begin{equation}
\begin{split}
&0<\frac{1}{2}c_2<\mathscr{X}<c_2, \\
&0<\frac{1}{2}\tilde{c}_2<\tilde{\mathscr{X}}<\tilde{c}_2. \\
\end{split}
\label{eq:para_kankeisiki}
\end{equation}

\begin{figure}
\includegraphics[width=90mm]{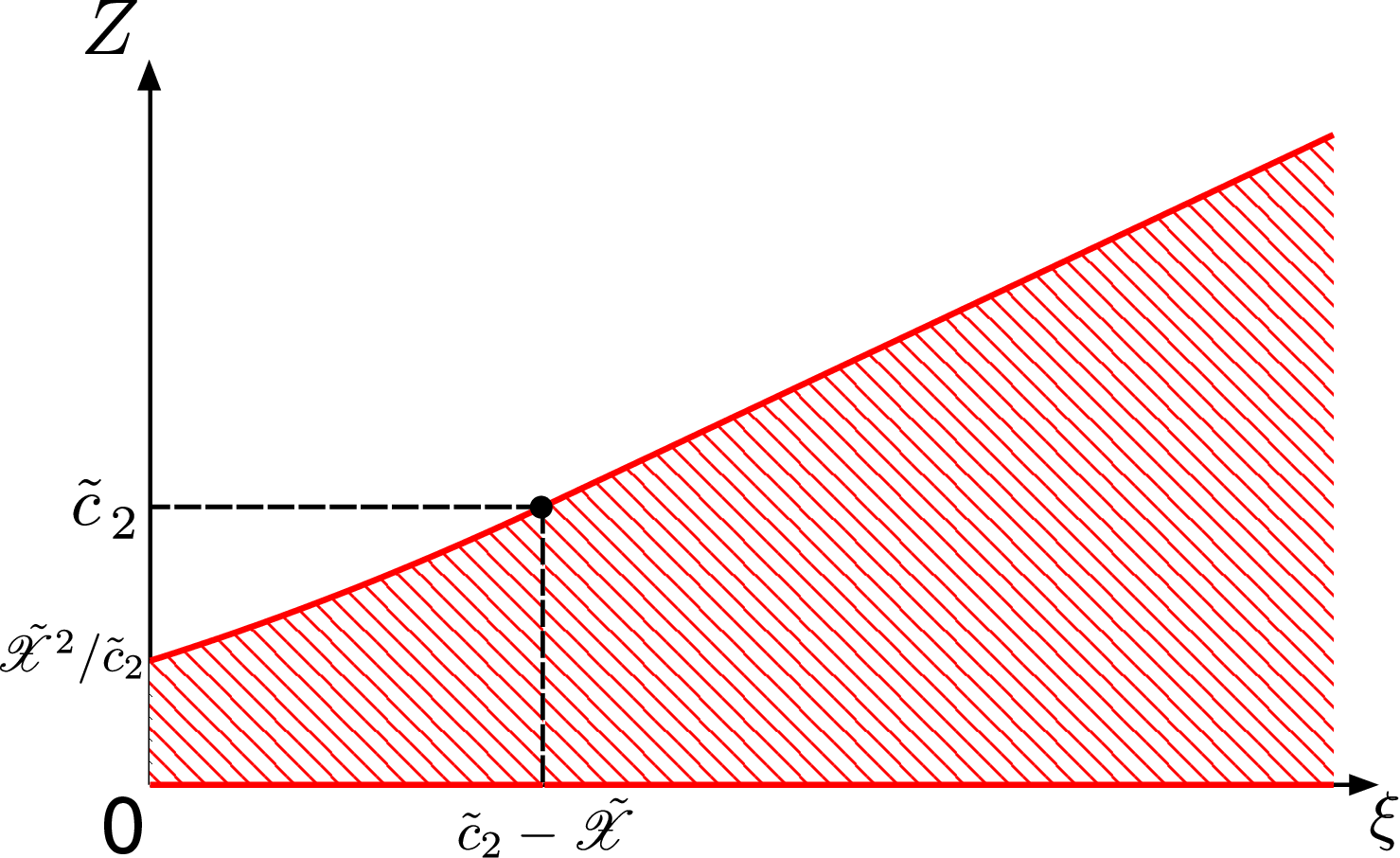}
\caption{
The possible range of $Z$ for each $\xi$ 
shown in 
Eq. (\ref{eq:xi_Z_c1>0}). 
\label{fig_1_xi_Z}}
\end{figure}

The possible range of $Z$ depends on the external magnetic 
flux density $B_s$ through $\xi$. 
For $c_1\ge 0$ and $j\ge3$, 
the range is given by 
\begin{equation}
\begin{cases}
\displaystyle
0\leq Z\leq\frac{1}{\tilde{c}_2}(\xi+\tilde{\mathscr{X}})^2\quad\quad\quad\quad\quad\quad\quad&\left(0\leq\xi\leq\tilde{c}_2-\tilde{\mathscr{X}}\right) \\
0\leq Z\leq2\xi+2\tilde{\mathscr{X}}-\tilde{c}_2 &\left(\tilde{c}_2-\tilde{\mathscr{X}}\leq\xi\right), \\
\end{cases}
\label{eq:xi_Z_c1>0}
\end{equation}
shown in Fig. \ref{fig_1_xi_Z}. 
The possible range for $c_1\leq0$ 
and $j\ge3$ is obtained by 
replacing $\tilde{c}_2$ and $\tilde{\mathscr{X}}$ 
by $c_2$ and $\mathscr{X}$, respectively. 
We suppose $c_1\geq0$ in the following.

\subsection{Equation for only $\alpha$}

Defining a quartic polynomial 
\begin{equation}
P(\alpha)=Q_1(\alpha)Q_2(\alpha)-c_1^2\frac{Y^2}{X^2}\alpha^2, 
\label{eq:P_teigi}
\end{equation}
the differential equation for only $\alpha$ is obtained: 
\begin{equation}\label{eq:bibun_alp}
\dot{\alpha}^2=P(\alpha). 
\end{equation}
Obviously, $\alpha$ moves in the 
region of $P(\alpha)\ge0$. 
Depending of the multiplicity of the 
roots of $P$, the behavior of $\alpha$ is classified into three categories. Namely, ``oscillates'', ``reflects at most once and converges'', and ``remains in the initial value''. 

The solution of Eq. (\ref{eq:bibun_alp}) is given as follows \cite{Shibata2022EPJD}. 
In the case of oscillation, $\alpha$ is a 
periodic function. 
Once the minimum and maximum values are 
determined, $\alpha$ can be described by 
the Jacobi's elliptic function $\text{sn}$. 
If $\alpha$ converges, 
$\dot{\alpha}$ changes its sign at most once. 
Hence, 
$\alpha$ can be obtained by integrating 
$\dot{\alpha}=\pm\sqrt{P}$ at most twice, 
by choosing a correct sign.

The remaining task is to specify the roots of $P$ 
which determine the behavior of $\alpha$. 
All the roots can be 
calculated by the quartic formula.
However, this method inevitably includes 
complex numbers, known as ``casus irreducibilis'' \cite{Galois}. 
It is hard to distinguish which root to be an upper or lower limit 
of $\alpha$. 
Thus, we treat the cases $Y=0$ and $Y\neq0$ separately.

\subsection{Behavior of $\alpha$ for $Y=0$}

In this case, 
$P(\alpha)=Q_1(\alpha)Q_2(\alpha)$. 
We notate the roots of 
$Q_1(\alpha)$ and $Q_2(\alpha)$ as 
$\mathfrak{q}_{1\pm}=\{ \xi+\mathscr{X} \pm
[(\xi+\mathscr{X})^2-c_2Z]^{1/2} \}/c_2$ and 
$\mathfrak{q}_{2\pm}=\{ \xi+\tilde{\mathscr{X}} \pm
[(\xi+\tilde{\mathscr{X}})^2-\tilde{c}_2Z]^{1/2} \}/\tilde{c}_2$, 
respectively. If the roots are double root, they are notated as 
$\mathfrak{q}_1$ and $\mathfrak{q}_2$, respectively.

The category of $\alpha$ is determined by $\xi$ and $Z$. 
For each category, characteristic values of $\alpha$ and 
corresponding ranges of $\xi$ and $Z$ are 
summarized in Tables \ref{table_a_shindou}, \ref{table_a_syuusoku}, and 
\ref{table_a_ugokanai}, respectively.

\begin{table}[H] 
\caption{The minimum and maximum values of 
oscillating $\alpha$ and corresponding ranges of 
$\xi$ and $Z$. 
The symbol $\land$ is a logical conjunction. 
\label{table_a_shindou}}
\begin{ruledtabular}
\begin{tabular}{cc}
Range of $\alpha$ 
& Ranges of $\xi$ and $Z$ \\
\hline
%%%%%%%%%%%%%%%%%%%%%%
$[\mathfrak{q}_{2-}, \mathfrak{q}_{1-}]$ & 
\begin{tabular}{c}
$0\leq\xi<c_2-\mathscr{X} \land 
0<Z<(\xi+\mathscr{X})^2/c_2$ or \\
%%%%%%%%%%%%%
$c_2-\mathscr{X}\leq \xi \land 
0<Z<2\xi+2\mathscr{X}-c_2
$ \\
\end{tabular} \\
%%%%%%%%%%%%%%%%%%%%%%
$[\mathfrak{q}_{1+}, \mathfrak{q}_{2+}]$ & 
$0\leq\xi<c_2-\mathscr{X} \land 
2\xi+2\mathscr{X}-c_2<Z<(\xi+\mathscr{X})^2/c_2
$ \\
%%%%%%%%%%%%%%%%%%%%%%
$[\mathfrak{q}_{2-}, \mathfrak{q}_{2+}]$ & 
\begin{tabular}{c}
$0\leq\xi<c_2-\mathscr{X} \land 
(\xi+\mathscr{X})^2/c_2<Z<
(\xi+\tilde{\mathscr{X}})^2/\tilde{c}_2
$ or \\
%%%%%%%%%%%%%
$c_2-\mathscr{X}\leq \xi<\tilde{c}_2-\tilde{\mathscr{X}} \land 
2\xi+2\mathscr{X}-c_2<Z<
(\xi+\tilde{\mathscr{X}})^2/\tilde{c}_2
$ \\
\end{tabular} \\
%%%%%%%%%%%%%%%%%%%%%%
\end{tabular}
\end{ruledtabular}
\end{table}

\begin{table}[H]
\caption{
The limit values 
of converging $\alpha$ and corresponding ranges of 
$\xi$ and $Z$. The initial value $\alpha(0)$ must differ 
from the limit value. 
\label{table_a_syuusoku}}
\begin{ruledtabular}
\begin{tabular}{cc}
Limit value of $\alpha$ 
& Ranges of $\xi$ and $Z$ \\
\hline
%%%%%%%%%%%%%%%%%%%%%%
$\mathfrak{q}_1$ & 
$0\leq\xi<c_2-\mathscr{X} \land 
Z=(\xi+\mathscr{X})^2/c_2$ \\
%%%%%%%%%%%%%%%%%%%%%%
$1$ & 
$ c_2-\mathscr{X}\leq\xi<\tilde{c}_2-\tilde{\mathscr{X}} \land 
Z=2\xi+2\mathscr{X}-c_2
$ \\
%%%%%%%%%%%%%%%%%%%%%%
\end{tabular}
\end{ruledtabular}
\end{table}

\begin{table}[H]
\caption{For stationary $\alpha$, its value and 
corresponding ranges of $\xi$ and $Z$. 
\label{table_a_ugokanai}}
\begin{ruledtabular}
\begin{tabular}{cc}
Stationary value of $\alpha$ 
& Ranges of $\xi$ and $Z$ \\
\hline
%%%%%%%%%%%%%%%%%%%%%%
$0$ & 
$0\leq\xi \land 
Z=0$ \\
%%%%%%%%%%%%%%%%%%%%%%
$\mathfrak{q}_1$ & 
$ 0\leq\xi<c_2-\mathscr{X} \land 
Z=(\xi+\mathscr{X})^2/c_2
$ \\
%%%%%%%%%%%%%%%%%%%%%%
$\mathfrak{q}_2$ & 
$ 0\leq\xi<\tilde{c}_2-\tilde{\mathscr{X}} \land 
Z=(\xi+\tilde{\mathscr{X}})^2/\tilde{c}_2
$ \\
%%%%%%%%%%%%%%%%%%%%%%
$1$ & 
$ 0\leq\xi \land 
Z=2\xi+2\mathscr{X}-c_2
$ \\
%%%%%%%%%%%%%%%%%%%%%%
\end{tabular}
\end{ruledtabular}
\end{table}

\subsection{Behavior of $\alpha$ for $Y\neq0$}

\begin{figure}[H]
\centering
\includegraphics[width=100mm]{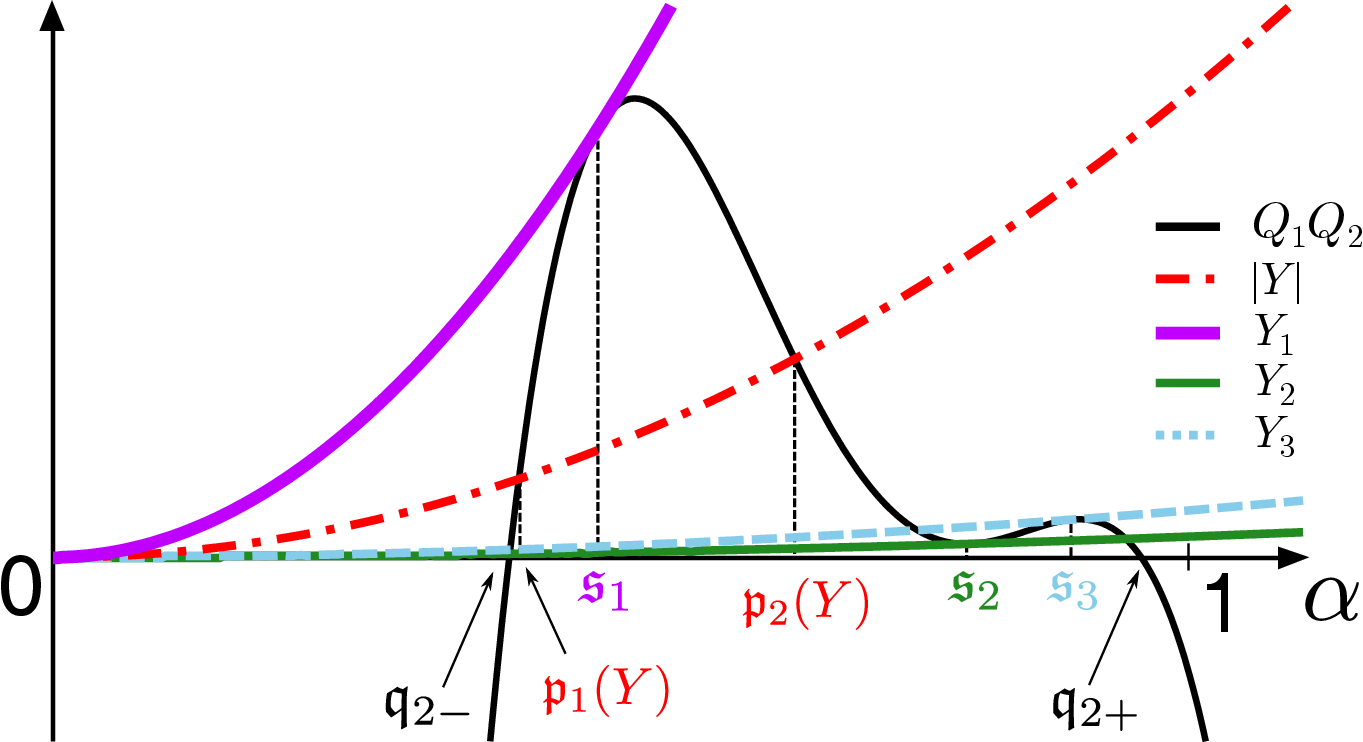}
\caption{
An example of $Q_1Q_2$ and $(c_1Y/X)^2\alpha^2$ 
for several $Y$. 
The parameters are set to 
$c_1/(C_{2,0}X)=2, 
c_2/(C_{2,0}X)=3, 
\mathscr{X}/(C_{2,0}X)=2, 
\xi/(C_{2,0}X)=0.5,$ and 
$Z/(C_{2,0}X)=2.1$ 
for the visibility. 
In this example, $Y_{max}=Y_1$ and 
$\tilde{\mathfrak{s}}=\mathfrak{s}_1$. 
\label{fig:kotei_Z}}
\end{figure}

$\alpha$ can move in the region of  
$Q_1Q_2\ge (c_1Y/X)^2\alpha^2$. 
Thus, $\alpha$ can be solved by 
obtaining the intersections of 
$Q_1Q_2$ and $(c_1Y/X)^2\alpha^2$. 
Figure \ref{fig:kotei_Z} shows an example for 
certain $c_1,c_2,\mathscr{X},\xi$ and $Z$. 
The black and red curves express $Q_1Q_2$ and 
$(c_1Y/X)^2\alpha^2$, respectively. 
$\alpha$ oscillates between 
$[\mathfrak{p}_1(Y), \mathfrak{p}_2(Y)]$, where 
$\mathfrak{p}_1(Y), \mathfrak{p}_2(Y), \dots$
be the roots of $P$, i.e.,  intersections of 
$Q_1Q_2$ and $(c_1Y/X)^2\alpha^2$ existing in 
$[\mathfrak{q}_{2-}, \mathfrak{q}_{2+}]$ 
in ascending order. 

Before explaining the figure in detail, 
we refer to several $Y$ such that 
$Q_1Q_2$ and $(c_1Y/X)^2\alpha^2$ are tangent in $[0,1]$. 
The corresponding $\alpha$ is a root of both $P$ and $P'$. 
Thus, it is a solution of the following 
equation: 
\begin{equation}\label{eq:bibun_setten}
\frac{Q_1'}{Q_1}+\frac{Q_2'}{Q_2}=\frac{2}{\alpha}, 
\end{equation}
which does not contain $Y$.
We express the solutions existing in $[0,1]$ 
in ascending order as 
$\mathfrak{s}_1, \mathfrak{s}_2, \dots$. 
The maximum number is three. 
For each solution, 
there is a unique $Y_{1,2,\dots}>0$ such that the 
solution to be a point of contact of 
$Q_1Q_2$ and $(c_1Y/X)^2\alpha^2$. 
For example, $\alpha=\mathfrak{s}_1$ 
is a solution of 
Eq. (\ref{eq:bibun_setten}) and $P(\mathfrak{s}_1)=0$ holds. 
The maximum value among $Y_{1,2,\dots}$ is 
the maximum value $Y_{max}$ of $Y$ and 
the corresponding solution of Eq. (\ref{eq:bibun_setten}) is 
notated by $\tilde{\mathfrak{s}}$.

We demonstrate 
how the value of $Y$ changes the behavior of $\alpha$ 
in Fig. \ref{fig:kotei_Z}. In this case, $Q_1$ has complex roots, 
$\alpha$ oscillates in 
$[\mathfrak{q}_{2-}, \mathfrak{q}_{2+}]$ at $Y=0$, 
and $Y_{max}=Y_1$.

The behavior of $\alpha$ is classified as follows. 
(i) If $0<|Y|<Y_2$, 
$\alpha$ oscillates between 
$[\mathfrak{p}_1(Y), \mathfrak{p}_2(Y)]$. 
%%%%%%%%%%%%%%%%%%%%%%%%%
(ii) If $|Y|=Y_2$, $\alpha$ converges to or remains in 
$\mathfrak{s}_2$. 
%%%%%%%%%%%%%%%%%%%%%%%%%
(iii) If $Y_2<|Y|<Y_3$, $\alpha$ oscillates between 
$[\mathfrak{p}_1(Y), \mathfrak{p}_2(Y)]$ 
or 
$[\mathfrak{p}_3(Y), \mathfrak{p}_4(Y)]$, 
depending on its initial value. 
%%%%%%%%%%%%%%%%%%%%%%%%%
(iv) If $|Y|=Y_3$, $\alpha$ oscillates between 
$[\mathfrak{p}_1(Y), \mathfrak{p}_2(Y)]$  or 
remains in $\mathfrak{s}_3$. 
%%%%%%%%%%%%%%%%%%%%%%%%%
(v) If $Y_3<|Y|<Y_1$, $\alpha$ oscillates between 
$[\mathfrak{p}_1(Y), \mathfrak{p}_2(Y)]$. 
This case is shown 
by the red chained curve. 
%%%%%%%%%%%%%%%%%%%%%%%%%
(vi) If $|Y|=Y_1$, $\alpha$ remains in 
$\tilde{\mathfrak{s}}=\mathfrak{s}_1$.


\end{document}